\documentclass[letterpaper,10pt]{article} 

\usepackage{opticameet3} 
\usepackage{physics}
\usepackage{amsmath}
\newcommand\authormark[1]{\textsuperscript{#1}}

\usepackage{amsmath,amssymb}
\usepackage[colorlinks=true,bookmarks=false,citecolor=blue,urlcolor=blue]{hyperref} 

\begin{document}

\title{  High-rate continuous-variable measurement-device-independent quantum key distribution}


\author{Adnan A.E. Hajomer,\authormark{*} Huy Q. Nguyen,\authormark{} and Tobias Gehring \authormark{**}}

\address{\authormark{} Center for Macroscopic Quantum States (bigQ), Department of Physics, Technical University of Denmark, 2800 Kongens Lyngby, Denmark}

\email{\authormark{*}aaeha@dtu.dk, \authormark{**} tobias.gehring@fysik.dtu.dk} 

\begin{abstract}
We report the first experiment of  continuous-variable measurement-device-independent quantum key distribution that enables secret key generation at a symbol rate of 5 MBaud without frequency and optical phase locking. This is achieved by  using a new relay structure  based on  a polarization-based 90-degree optical hybrid and a well-designed DSP pipeline.     
\end{abstract}

\section{Introduction}
Quantum key distribution(QKD) is a secure means of distributing cryptographic keys based on the laws of quantum physics~\cite{stefano}. While QKD offers information-theoretic security in theory, the physical devices have practical imperfections which  can result in side channels that can be exploited by an eavesdropper. For instance, various attacks on detectors have been reported as these are the most vulnerable part of QKD systems~\cite{HKLC}. 

To cope with these side channels,  measurement-device-independent QKD (MDI-QKD) has been proposed~\cite{S. L. Braunstein, HKLC2}. Here, the two communicating users (Alice and Bob) are connected by an untrusted third party, which performs a Bell-state measurement (BSM). This measurement acts as a correlator between Alice and Bob but does not reveal their symbol values. Thereby, MDI-QKD can eliminate all detector side-channel attacks. 

A continuous variable (CV) MDI-QKD was first proposed~\cite{stefano2}, independently reproposed~\cite{Z. Li}, and experimentally demonstrated ~\cite{stefano2,Y. Tian}  achieving a higher secret key rate (per channel use) compared with its discrete variable counterpart. These demonstrations, however, required a complex frequency and optical phase locking system to perform CV-BSM, making the practical realization of CV-MDI-QKD quite challenging. Furthermore, both experiments used a quantum symbol rate of a few hundred kHz, limiting the achievable secret key rate. 

Here, we report the first experimental CV-MDI-QKD system operating at a symbol rate of 5 Mbaud and without frequency and optical phase locking. Leveraging the concept of the polarization-based 90-degree hybrid~\cite{L. G. Kazovsky}, we develop a new relay structure, that enables the realization of CV-BSM  without phase locking. Simultaneously, our system uses a continuous-wave laser with digital pulse shaping and digital time synchronization, and therefore, it does not require additional amplitude modulation for pulse carving or a delay line for time synchronization. Combining these technologies with quadrature remapping~\cite{Y. Tian}, we realize  a practical and simple CV-MDI-QKD system, achieving a secret key rate of 0.12 bit per relay use, or correspondingly 600 kbit per second, over 2 dB loss corresponding to  10 km optical fiber channel (at the loss of 0.2 dB/km).  

\section{Basic concept and system}
In the prepare-and-measure (PM) scheme of CV-MDI QKD protocol, Alice and Bob prepare coherent states $\ket{\alpha}$ and $\ket{\beta}$, respectively, whose amplitudes $\alpha$ and $\beta$ are randomly drawn from a Gaussian distribution with zero mean and sufficiently large modulation variance in  each quadrature. Then, Alice and Bob send their coherent states to the intermediate station (Relay) through two independent quantum channels, where a CV-BSM is performed by mixing the incoming signals at a balanced beam splitter followed by double homodyne detection. The output of the measurement ($\gamma$) is then sent to Alice and Bob via a classical public channel.  With the knowledge of $\gamma$, either Alice or Bob infers the variable of the other party. Finally, Alice and Bob perform classical processing including parameter estimation, information reconciliation, and privacy amplification.

\begin{figure}[htbp]
  \centering
  \includegraphics[width=\linewidth]{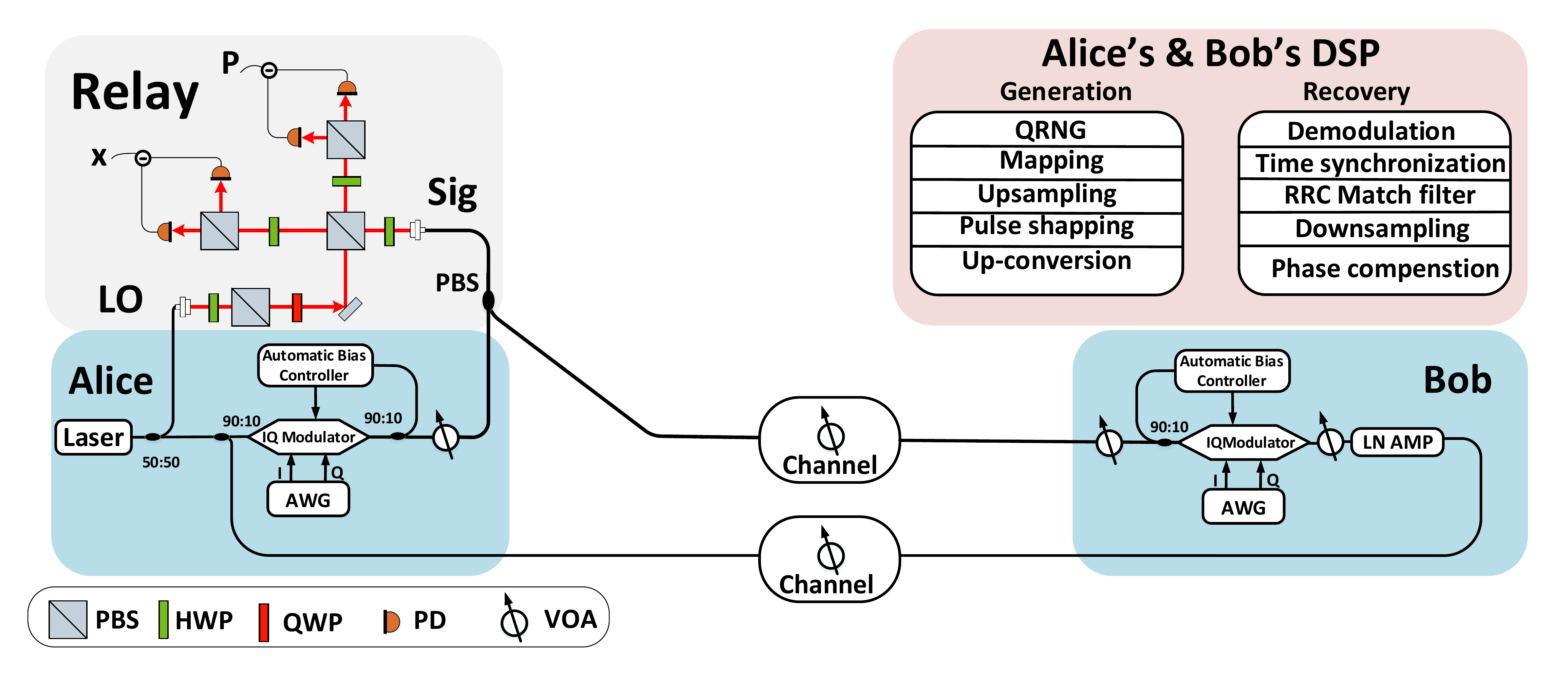}
\caption{CV-MDI-QKD system experimental setup and DPS pipeline. AWG:  arbitrary waveform generator; PBS: polarization beam splitter; HWP: Half-wave plate; QWP: quarter-wave plate; PD: photodiode; VOA: variable optical attenuator. }
\label{fig:exp}
\end{figure}

Figure 1 shows the schematic of our CV-MDI-QKD system. It consists of an optical layout and digital signal processing (DSP) pipeline for waveform generation and quantum symbols recovery. To prepare random coherent states, Alice and Bob drew random numbers at a rate of 5 MBaud from Gaussian distributions obtained by transforming the output of a vacuum-based quantum random number generator (QRNG)~\cite{T. Gehring}, forming the complex amplitude $\alpha$ and $\beta$ of their coherent states. The quantum symbols were then upsampled to 1 GSample/s and pulse-shaped using a root-raised cosine filter with a roll-off of 0.2. To avoid the low-frequency noise of the transmitter, the quantum signal was digitally up-converted to $\omega/2\pi=$5 MHz, i.e.\ multiplied with $\cos(\omega t)$, for double sideband modulation. Finally, Alice and Bob uploaded their waveforms to dual-channel arbitrary waveform generators (AWGs) each with 16-bit resolution and a sampling rate of 1 GSample/s. 

  The transmitters of Alice and Bob were built from polarization maintaining fiber components. At Alice's station, a 1550 nm continuous-wave laser with a linewidth of $\approx$ 100 Hz was shared with the relay, as the optimal configuration of the CV-MDI protocol is asymmetric with the relay being at Alice~\cite{stefano2}. To avoid frequency locking, a portion of Alice's laser of 1.9 mW was also sent to Bob through an independent fiber channel emulated by a variable optical attenuator (VOA) with a loss of 2 dB (corresponding to a 10 km fiber channel) and then amplified at Bob's station using a low noise amplifier (LN AMP). In each transmitter, an in-phase and quadrature (IQ) modulator driven by the AWG was  used to prepare the ensemble of coherent states. The DC biases to these IQ modulators were controlled using commercial automatic bias controllers. To adjust the modulation variance of the coherent states two VOAs were used at Alice's and Bob's sides. At Bob's side the VOA was also used to simulate the fiber channel to the relay. As the relay is situated at Alice's station our implementation thus only requires a commonly available fiber pair.

At the relay, the incoming beams from Alice and Bob were overlapped at a fiber-based polarization beamsplitter (PBS). The signal was then free-space coupled into the polarization-based 90-degree hybrid, where  the two  quadratures (X and P)  were detected simultaneously. This is possible because the local oscillator (LO) was prepared in circular polarization by means of a quarter wave-plate (QWP), while the signal was linearly polarized ~\cite{L. G. Kazovsky}. The hybrid was built from free-space bulk components with negligible losses to achieve high-efficiency CV-BSM. Two custom-made balanced detectors each with a bandwidth of 10 MHz and quantum efficiency of $99\%$ were used. Considering the insertion loss and the visibility of the interference fringes, the total detection efficiency of the relay was 94 $\%$. The output of the balanced detector was then digitized at a sampling rate of 1 GSample/s using an analog-to-digital converter (ADC), which was clock synchronized to Alice’s and Bob's AWG using a 10 MHz external reference. The measurement time was divided into frames, each with $10^7$ samples. Furthermore, the one-time shot noise calibration technique was used for system calibration~\cite{Y. Zhang}.

After the relay broadcasting the output of CV-BSM, Alice's and Bob's DSP algorithms for quantum symbols recovery were applied. The steps are shown in Fig.~\ref{fig:exp}. First, the relay output was downconverted to the baseband  and low pass filtered. Temporal synchronization was achieved through the cross-correlation between Alice's and Bob's transmitted samples and the relay output. In contrast to one-way CVQKD, the propagation delay was compensated on the transmitted samples since each quantum channel had a different propagation delay. The synchronized samples were then matched filtered and downsampled to symbols. As one laser was shared between the communicating parties and the relay, Alice's and Bob's signals experienced a slow phase drift with respect to the LO. To compensate for this phase drift, Alice and Bob rotated their symbols to maximize the correlation as: 
$\hat{\alpha} =  \underset{\theta_1} {\arg\max}~Cov(\alpha~\exp(j\theta_1), \gamma), ~\hat{\beta} =  \underset{\theta_2} {\arg\max}~Cov(\beta~\exp(j\theta_2), \gamma)$,
where $Cov(.)$ denotes the covariance. Finally,  Alice and Bob performed displacement operations~\cite{stefano2} to correlate their symbols according to the CV-BSM output.  

\begin{figure}[t]
  \centering
  \includegraphics[width=1.1\linewidth]{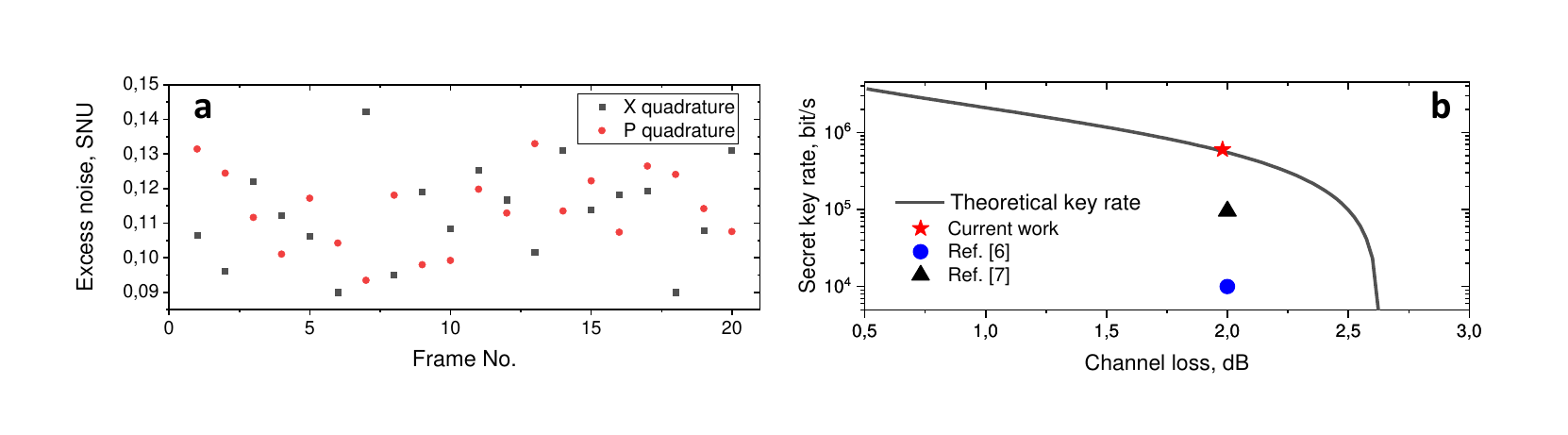}
  \vspace{-1.1cm}
\caption{ Experimental results demonstrating our CV-MDI-QKD system performance. (a) Excess noise variance (in shot-noise units). (b) Experimental key rates and numerical simulations. }
\label{fig:result}
\end{figure}

\begin{table}[t]
 \centering \vspace{-0.5cm}\caption{Experimental parameters}
\begin{tabular}{ccccccc}
    \hline
   symbol rate&~~~~$V_\text{Alice}$ & $V_\text{Bob}$ & $\xi$ & $\eta$ & $\tau_\text{Bob}$\\
    \hline
    5 Mbaud &~~~~36 SNU & 36 SNU & 0.11 SNU  & 0.94  \%  & 2 dB\\
    \hline
   \end{tabular}
   \label{fig:tab1}
    \end{table}

\section{Results}
Table~\ref{fig:tab1} summarizes the relevant parameters in our experiment. Alice and Bob  prepared an ensemble of $4\times10^6$ coherent states, characterized by  the modulation variance of $V_\text{Alice}$ and $ V_\text{Bob}$. As the modulation variance directly affects the excess noise $\xi$, our system operates at a modulation variance of 36 shot-noise units (SNU) to reduce the excess noise. Fig.~\ref{fig:result}(a) depicts the measured excess noise variance for 20 frames each with $2 \times 10^5$ symbols. The average excess noise of X and P quadratures is 0.11 SNU. The corresponding secret key rate is evaluated in the asymptotic regime according to~\cite{stefano2}. Fig.~\ref{fig:result}(b) shows the secret key rate of the numerical simulation and experimental results. The black solid curve shows the numerical simulations calculated from experimental parameters. Considering the information reconciliation of $97\%$ efficiency and Bob's channel loss $\tau_{Bob}$ of 2 dB, we achieved a secret key fraction of 0.12 bit per relay use which corresponds to 600 kbit secret key per second.
\section{Discussion}
In this work, we demonstrated the first simple and practical CV-MDI-QKD system, which is free of frequency and optical phase locking and operates at a 5 Mbaud symbol rate. This was enabled by means of a new relay structure leveraging  the concept of a polarization-based 90-degree optical hybrid and DSP for CV-BSM. Compared to the previous demonstration~\cite{stefano2, Y. Tian}, our system improves the secret key rate by one order of magnitude, which boosts the secret key rate to 0.6 Mbps. This is six times higher than the record secret key bit per second \cite{Y. Tian}. However, the symbols rate can be further increased using a broadband balanced detector. Our demonstration could pave the way toward the practical adoption of CV-MDI-QKD to build high-rate quantum networks.  
\vspace{0.5cm}

\noindent \paragraph{Acknowledgments} This work was supported by the European Union’s Horizon 2020 research and innovation programs through CiViQ (grant agreement no. 820466) and OPENQKD (grant agreement no.\ 857156) and from the Danish National Research Foundation, Center for Macroscopic Quantum States (bigQ, DNRF142).

\end{document}